\documentclass{article}
\usepackage{spconf,amsmath,graphicx}

\usepackage{amsmath}
\usepackage{amssymb}
\usepackage{pgfplotstable}
\pgfplotsset{compat=1.18}
\usepackage{booktabs}
\usepackage{adjustbox}
\usepackage{multirow}
\usepackage{hyperref}
\usepackage{booktabs,colortbl}
\usepackage{pifont} 

\title{Deep Generic Representations for Domain-Generalized Anomalous Sound Detection}
\name{Phurich Saengthong, Takahiro Shinozaki}
\address{Tokyo Institute of Technology \\ \small \texttt{saengthong.p.aa@m.titech.ac.jp}}
\begin{document}
%
\maketitle
\begin{abstract}
Developing a reliable anomalous sound detection (ASD) system requires robustness to noise, adaptation to domain shifts, and effective performance with limited training data. Current leading methods rely on extensive labeled data for each target machine type to train feature extractors using Outlier-Exposure (OE) techniques, yet their performance on the target domain remains sub-optimal. In this paper, we present \textit{GenRep}, which utilizes generic feature representations from a robust, large-scale pre-trained feature extractor combined with kNN for domain-generalized ASD, without the need for fine-tuning. \textit{GenRep} incorporates MemMixup, a simple approach for augmenting the target memory bank using nearest source samples, paired with a domain normalization technique to address the imbalance between source and target domains. \textit{GenRep} outperforms the best OE-based approach without a need for labeled data with an Official Score of 73.79\% on the DCASE2023T2 Eval set and demonstrates robustness under limited data scenarios. The code is available open-source \footnote{\url{https://github.com/Phuriches/GenRepASD}}.
\end{abstract}
\section{Introduction}
\label{sec:intro}
Developing a reliable anomalous sound detection (ASD) system for machine condition monitoring requires robustness to noisy inputs, adaptability to domain shifts (e.g., changes in machine states due to temperature or background noise), and effective performance with limited training data from new installations \cite{dohi_description_2023}. Key ASD approaches include unsupervised methods using generative models, such as reconstruction-based methods with autoencoders (AE) \cite{koizumi_description_2020}\cite{suefusa_anomalous_2020}\cite{wichern_anomalous_2021}\cite{10095568}\cite{10447941}, and distribution-based methods using auto-regressive models to predict probability density functions \cite{lee2021robust}. Self-supervised methods often assume labeled data for training a model and typically involve classification-based methods \cite{liu_anomalous_2022full}\cite{wilkinghoff_self-supervised_2023_full}.

Classification-based methods often achieve the best performance, with a focus on enhancing feature extractors via Outlier-Exposure (OE) techniques \cite{hendrycks2019oe}\cite{liu_anomalous_2022full}\cite{wilkinghoff_self-supervised_2023_full}. These techniques involve training on auxiliary tasks that distinguish between target classes based on machine types and states using cross-entropy loss, allowing the model to learn robust representations of target distributions. Anomaly detection is then performed by assigning classification scores \cite{liu_anomalous_2022full} or measuring distances between feature embeddings of known normal and unknown inputs from the pre-trained extractor \cite{wilkinghoff_self-supervised_2023_full}, which can also be considered as a distance-based method.

However, these methods face two key challenges \cite{dohi_description_2023}\cite{koizumi_description_2020}: (1) they require sufficient normal data for training on both source and target domains, along with careful hyperparameter tuning to prevent overfitting and the capture of irrelevant noise; and (2) they necessitate extensive labeling for machine types and section IDs on normal and anomalous data, which can be impractical or infeasible in real-world applications. This raises the quesiton: How can robust feature representations be obtained without these challenges?

Recent advances in large-scale pre-trained speech and audio models have shown success in general audio tasks by learning robust, transferable features. To address domain-generalized ASD challenges, we introduce \textit{GenRep}, which leverages generic representations from BEATs \cite{pmlr-v202-chen23ag} combined with a distance-based approach, k-nearest neighbors (kNN). \textit{GenRep} incorporates MemMixup to balance source and target memory banks by augmenting target features with their nearest source features, and applies Domain Normalization (DN) using Z-score normalization to standardize score distributions across domains. 

Our contributions include: (1) demonstrating the effectiveness of robust, general-purpose feature representations using BEATs \cite{pmlr-v202-chen23ag} without fine-tuning for ASD; (2) evaluating performance across BEATs' pre-training methods and layers, proposing an effective pooling strategy, and mitigating domain shift with introduced MemMixup and DN; (3) achieving state-of-the-art performance on domain-shift settings with an Official Score of 73.79\% on the DCASE2023T2 Eval set \cite{Harada2021}, while also demonstrating robustness under limited data scenarios on the DCASE2020T2 Dev set \cite{Koizumi_WASPAA2019_01}\cite{Purohit_DCASE2019_01}.

\noindent{\textbf{Related work.}}
Our work relates to both reconstruction-based and classification-based anomaly detection methods. Reconstruction-based approaches, like autoencoders (AE) \cite{koizumi_description_2020}\cite{suefusa_anomalous_2020}\cite{wichern_anomalous_2021}\cite{10095568}\cite{10447941}, and kNN \cite{eskin_geometric_2002} estimate anomaly scores by comparing test samples to reconstructed normal data. Classification-based methods typically involve training feature extractors with techniques like Outlier-Exposure (OE) \cite{hendrycks2019oe} and then applying anomaly detection on the learned representations \cite{wilkinghoff_self-supervised_2023_full}\cite{dohi_description_2023}. In contrast, our approach uses a pre-trained audio feature extractor, BEATs \cite{pmlr-v202-chen23ag}, with kNN, without the need for additional fine-tuning. Our work also related to image anomaly detection methods that use pre-trained feature extractors without fine-tuning \cite{bergman_deep_2020}, but utilize audio Transformers \cite{NIPS2017_3f5ee243} pre-trained on AudioSet \cite{gemmeke_audio_2017}, which may affect feature extraction effectiveness for ASD and domain-generalized tasks, differently.

\section{Method}
\label{sec:pagestyle}
\subsection{Feature extraction and pooling}
\label{ssec:fe}
To obtain feature embeddings for a memory bank, \textit{GenRep} utilizes a pre-trained model on AudioSet, such as BEATs \cite{pmlr-v202-chen23ag}, which is based on the ViT \cite{dosovitskiy2021an} architecture with multiple Transformer \cite{NIPS2017_3f5ee243} layers. The pre-trained model choice is flexible, allowing for unsupervised, semi-supervised, or supervised approaches \cite{pmlr-v202-chen23ag}. Given a set of input spectrograms \( X_{\text{train}} = \{ x_1, x_2, \ldots, x_N \} \), the model maps each input to a patch embedding \( x_i \in \mathbb{R}^{T \times F \times C} \), where \( T \), \( F \), and \( C \) represent time frames, frequency bins, and channels, respectively. The temporal and spectral dimensions are then flattened into a sequence \( x_i \in \mathbb{R}^{TF \times C} \) and processed by Transformer layers to produce embeddings \( f_i = F(x_i) \) with shape \( f_i \in \mathbb{R}^{TF \times C} \). While pooling over the entire spatial dimension $TF$ \cite{pmlr-v202-chen23ag} is an option, it can lead to sub-optimal performance as spectral components are not directly used in kNN-based anomaly scoring. Instead, we apply mean-pooling over the temporal dimension only, preserving spectral components by reshaping the embedding to \( f_i \in \mathbb{R}^{T \times F \times C} \) and then applying:
\begin{equation}
f_i = \frac{1}{T} \sum_{t=1}^{T} f_{i,t}.
\end{equation}
This results in a final embedding flattened into \( f_i \in \mathbb{R}^{FC} \).

\subsection{Memory Mixing up for handling domain shift}
When the target feature set $F_T$ is much smaller than the source set $F_S$, the proximity of source features to target test samples can introduce bias in anomaly detection performance. To address this imbalance, we propose MemMixup, a method that augments target features in the target memory bank by interpolating between each target feature and its nearest K source features. Unlike Mixup \cite{zhang2018mixup}, which augments samples during training to enhance overall model robustness, MemMixup specifically aims to increase the target features in a target memory bank by generating new features that lie between the source and target distributions, with a stronger emphasis on the target distribution. Specifically, we use kNN to find the nearest K source features for each target feature:
\begin{equation}
    d(f_t, f_s) = \|f_s - f_t\|^2,
\end{equation}
where $d(f_t, f_s)$ denotes the Euclidean distance between target feature $f_t$ and source feature $f_s$. We identify the top $K_s$ nearest source features to each target feature $f_t$, denoted as $\text{topk}(f_t) = \{f_s^{(1)}, f_s^{(2)}, \ldots, f_s^{(K_s)}\}$, where $\text{topk}(f_t)$ indicates the indices of these closest $K_s$ source features. Using these indices, we interpolate between each target feature and its nearest source features:
\begin{equation}
    \tilde{f_t}^{(i)} = \lambda f_t + (1 - \lambda) f_s^{(i)}, \quad \forall f_s^{(i)} \in \text{topk}(f_t),
\end{equation}
where $\lambda \in [0, 1]$ and is explicitly set to 0.9. This operation augments the set of target features in the memory bank, improving the balance between source and target samples. First, we define the set of augmented target features as:
\[
\tilde{F}_T = \{ \tilde{f}_{t_1}^{(1)}, \tilde{f}_{t_1}^{(2)}, \ldots, \tilde{f}_{t_1}^{(K_s)}, \ldots, \tilde{f}_{t_N}^{(1)}, \tilde{f}_{t_N}^{(2)}, \ldots, \tilde{f}_{t_N}^{(K_s)} \}.
\]
Then, the target memory bank stores the enriched set, denoted as $F_{\text{aug}} = F_T \cup \tilde{F}_T$, where $F_{\text{aug}}$ includes both the original target features $F_T$ and the augmented target features $\tilde{F}_T$.

\subsection{Anomaly detection}
\label{ssec:anomalydetection}
Using the features in the source and target memory banks, we determine if a test sample \( y \) is anomalous by computing its kNN distance separately from both the source and target memory banks. The anomaly score \( d(y) \) is defined as:
\begin{equation}
    d(y) = \frac{1}{K_n} \sum_{f \in N_{K_n}(f_y)} \|f - f_y\|^2,
\end{equation}
where \( f \) represents either the source features \( f_s \) or the target features \( f_t \), depending on whether we are calculating the distance from the source or target memory bank. Specifically, \( N_{K_n}(f_y) \) denotes the \( K_n \) nearest normal features \( f \) to the test sample \( f_y \) within the corresponding memory bank.

In domain shift scenarios, aligning source and target distances is essential to prevent performance degradation due to score discrepancies. To address this, we apply DN by separating source and target scores from their respective memory banks and normalizing them using Z-score. The Z-scores are computed as \(\text{Z-score}(d_s) = \frac{d_s - \mu_s}{\sigma_s}\) and \(\text{Z-score}(d_t) = \frac{d_t - \mu_t}{\sigma_t}\), where \(\mu_s, \sigma_s\) and \(\mu_t, \sigma_t\) are the means and standard deviations of the source and target scores, respectively. We then determine the closest domain sample for each test sample by finding the minimum distance between the normalized scores from both domains:
\begin{equation}
    \text{score} = \arg\min \left(\text{Z-score}(d_s(y)), \text{Z-score}(d_t(y))\right).
\end{equation}
This approach aims to align the score distributions, which may help reduce the impact of domain shift and potentially improve the overall performance of ASD task.

\newcommand{\highestvalue}[1]{\textbf{#1}}
\newcommand{\secondhighestvalue}[1]{\underline{#1}}
\pgfplotstableread[col sep=comma]{data/table1_new3.csv}\datatable

\begin{table*}[t]
\centering
\caption{Domain-generalized anomalous sound detection performance on DCASE2023T2 Eval set \cite{dohi_description_2023}.}
\label{tab:result_official}
\begin{adjustbox}{width=0.97\textwidth}
\large
\pgfplotstabletypeset[
  string type,
  columns/Trainingsetting/.style={string type, column name=Training Setting, column type=c},
  columns/Anomalydetection/.style={string type, column name=Anomaly detection, column type=c},
  columns/AUCsource/.style={column name=AUC Source, column type=c, precision=2, fixed zerofill},
  columns/AUCtarget/.style={column name=AUC Target, column type=c, precision=2, fixed zerofill},
  columns/pAUC/.style={column name={pAUC}, column type=c, precision=2, fixed zerofill},
  columns/OC/.style={column name={Official Score}, column type=c, precision=2, fixed zerofill},
  every head row/.style={
    before row={\toprule
      & \multicolumn{4}{c}{DCASE2023T2 \cite{dohi_description_2023}} & \multicolumn{1}{c}{OE} & \multicolumn{3}{c}{Ours}\\
    },
    after row=\midrule,
  },
  every col no 0/.style={string type, column type=l, column name=Method},
  every row no 6/.style={after row=\midrule},
  every row no 8/.style={after row=\midrule},
  columns/AE/.style={column name={AE}},
  columns/Ranked 3rd/.style={column name={Ranked 3rd}},
  columns/Ranked 2nd/.style={column name={Ranked 2nd}},
  columns/Ranked 1st/.style={column name={Ranked 1st}},
  columns/SSLASD/.style={column name={SSL4ASD \cite{wilkinghoff_self-supervised_2023_full}}},
  columns/GenRep100/.style={column name={GenRep\textendash 100}},
  columns/GenRep500/.style={column name={GenRep\textendash 500}},
  columns/GenRep990/.style={column name={GenRep\textendash 990}},
  every last row/.style={after row=\bottomrule},
  every row 0 column 8/.style={postproc cell content/.style={@cell content=\highestvalue{72.71}/56.58}},
  every row 0 column 6/.style={postproc cell content/.style={@cell content=71.96/\highestvalue{57.63}}},
  every row 1 column 4/.style={postproc cell content/.style={@cell content=\highestvalue{89.03}/\highestvalue{77.74}}},
  every row 2 column 3/.style={postproc cell content/.style={@cell content=\highestvalue{74.80}/\highestvalue{63.79}}},
  every row 3 column 6/.style={postproc cell content/.style={@cell content=\highestvalue{96.82}/\highestvalue{87.00}}},
  every row 4 column 6/.style={postproc cell content/.style={@cell content=78.85/\highestvalue{58.07}}},
  every row 4 column 8/.style={postproc cell content/.style={@cell content=\highestvalue{79.02}/57.76}},
  every row 5 column 8/.style={postproc cell content/.style={@cell content=\highestvalue{77.28}/\highestvalue{62.67}}},
  every row 6 column 6/.style={postproc cell content/.style={@cell content=84.33/\highestvalue{77.76}}},
  every row 6 column 8/.style={postproc cell content/.style={@cell content=\highestvalue{84.37}/77.66}},
  every row 7 column 7/.style={postproc cell content/.style={@cell content=79.50/\highestvalue{64.83}}},
  every row 7 column 8/.style={postproc cell content/.style={@cell content=\highestvalue{79.67}/64.71}},
  every row 8 column 4/.style={postproc cell content/.style={@cell content=\highestvalue{83.13}/60.08}},
  every row 8 column 8/.style={postproc cell content/.style={@cell content=81.30/\highestvalue{77.51}}},
  every row 9 column 8/.style={postproc cell content/.style={@cell content=\highestvalue{73.79}}},
]\datatable
\end{adjustbox}
\end{table*}

\section{Experiments}
\label{sec:majhead}

\begin{figure}
    \begin{minipage}[b]{1.0\linewidth}
      \centering
      \centerline{\includegraphics[width=8.5cm]{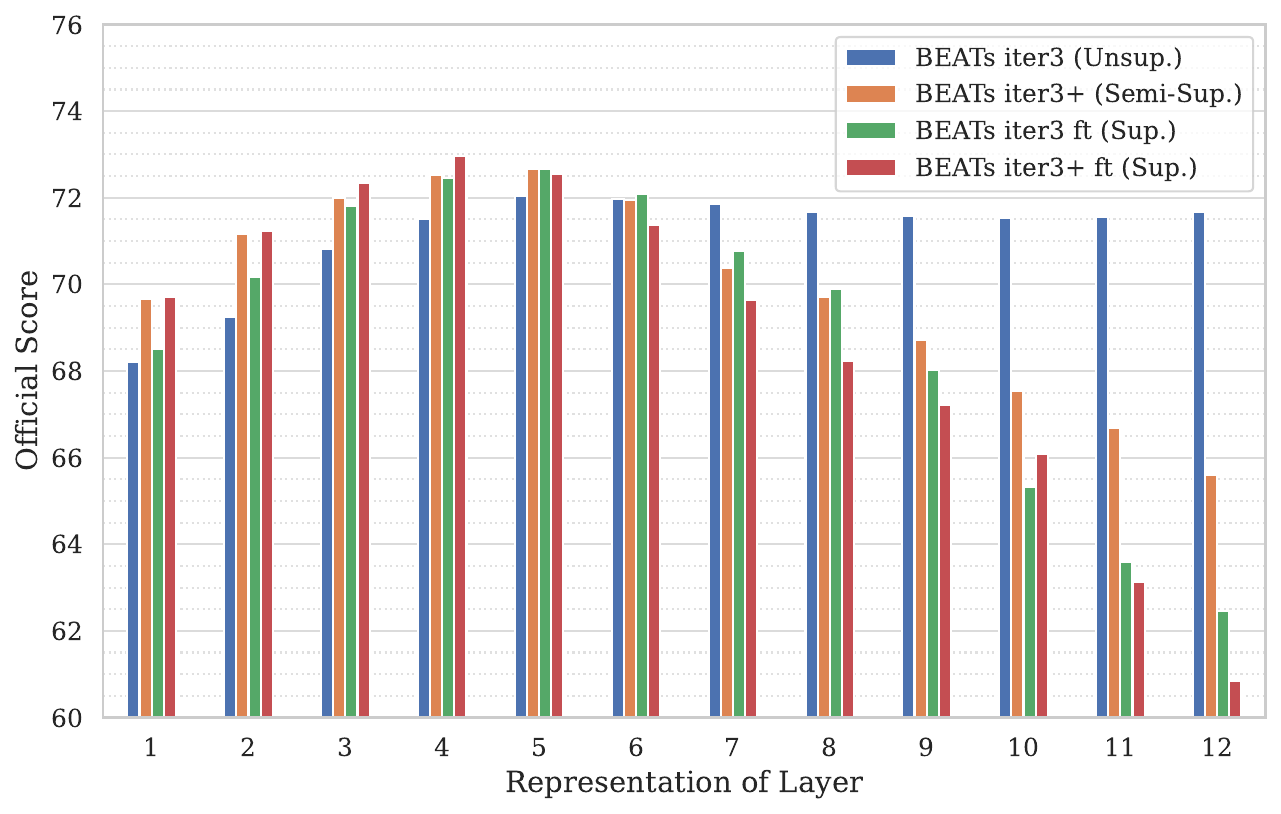}}
    \end{minipage}
    \caption{Comparison of performances of \textit{GenRep} without applying MemMixup using each BEATs\cite{pmlr-v202-chen23ag}' layer on DCASE2023T2 Eval set \cite{dohi_description_2023}.}
    \label{fig:layereffect}
\end{figure}

\subsection{Setup}
\textbf{Datasets and Metrics.} We use the DCASE2023T2 Eval set for domain-shift settings and the DCASE2020T2 Dev set for low-shot settings. Both datasets include machine sounds recorded in labs and mixed with real factory noise. The DCASE2023T2 Eval set \cite{dohi_description_2023} is designed for domain-generalized ASD, with 990 source and 10 target samples per machine type, featuring sounds from seven ToyAdmos2+ machines \cite{haradatoyadmos2+}. Recordings are single-channel, 6-18 seconds long at 16 kHz. The DCASE2020T2 Dev set \cite{koizumi_description_2020} includes sounds from six machine types sourced from ToyADMOS \cite{Koizumi_WASPAA2019_01} and MIMII \cite{Purohit_DCASE2019_01} datasets, with 10-second recordings at 16 kHz. For domain-shift, results are based on AUC, partial AUC (pAUC), and the Official Score using the harmonic mean of source AUC, target AUC, and mixed pAUC across all machine types \cite{dohi_description_2023}. For low-shot, results are reported as the arithmetic mean of AUC and pAUC scores across all machine types.

\textbf{Implementation details.} We use the default BEATs \cite{pmlr-v202-chen23ag} implementation for input feature extraction. Each raw waveform is sampled at 16 kHz with a duration of 10 seconds and converted into 128-dimensional Mel-filter bank features using a 25 ms Povey window with a 10 ms shift. The features are normalized using the mean and standard deviation values provided by \cite{pmlr-v202-chen23ag}. Each feature is split into 16 × 16 patches and flattened into a sequence for input. For augmentation, we apply MemMixup to each target feature with $\lambda$ set to 0.9. For the anomaly detection model, we set $K_n = 1$ for kNN. Results are reported based on the representation layers that yield the best performance for each machine type, unless otherwise specified.

\subsection{Robustness to domain shift}
\label{sssec:domainshift}
\subsubsection{Effect of feature representation in each layer}
\label{sssec:featureeffect}
Figure \ref{fig:layereffect} compares the performance of each layer’s representation without applying MemMixup from $\text{BEATs}{\substack{\text{iter3}}}$, $\text{BEATs}{\substack{\text{iter3+}}}$, and their fine-tuned versions, corresponding to unsupervised, semi-supervised, and supervised models, respectively. First, we observe that the best performances of the supervised fine-tuned models are higher than those of the unsupervised and semi-supervised pre-trained models. Second, the performance of the semi-supervised and supervised models declines from layer 6 to layer 12, while it remains stable for the unsupervised model. This suggests that the features from the deeper layers, particularly the last layer, of models trained with labels from AudioSet classes may be biased towards those classes. Finally, we find that layer 4 is the optimal layer for the supervised $\text{BEATs}{\substack{\text{iter3+}}}$, achieving an Official Score of 72.96, whereas layer 5 is optimal for the other models, with official scores exceeding 72. Based on these results, we will focus on utilizing the supervised $\text{BEATs}{\substack{\text{iter3+}}}$ (referred to as $\text{BEATs}_{\substack{\text{iter3+}}}$ ft), as it consistently demonstrates the best performance.

\newcommand{\cmark}{\ding{51}} 
\newcommand{\xmark}{\ding{55}} 

\pgfplotstableread[col sep=comma]{data/table4_new2.csv}\datatable
\begin{table}[t]
\centering
\caption{Comparison of applying domain score normalization and MemMixup.}
\label{tab:domainnorm}
\begin{adjustbox}{width=0.48\textwidth}
\pgfplotstabletypeset[
  col sep=comma,
  string type,
  columns/Model/.style={string type, column type=l, column name={Model}},
  columns/Mixup/.style={string type, column name=MemMixup, column type=c},
  columns/AUCsource/.style={column name=Source, column type=c, precision=2, fixed zerofill},
  columns/AUCtarget/.style={column name=Target, column type=c, precision=2, fixed zerofill},
  columns/OC/.style={column name=Official Score, column type=c, precision=2, fixed zerofill},
  every col no 0/.style={string type, column type=l, column name=Model},
  every row 0 column 1/.style={string replace={no}{\xmark}},
  every row 1 column 1/.style={string replace={no}{\xmark}},
  every row 0 column 2/.style={string replace={no}{\xmark}},
  every row 1 column 2/.style={string replace={no}{\xmark}},
  every row 2 column 1/.style={string replace={no}{\xmark}},
  every row 3 column 1/.style={string replace={no}{\xmark}},
  every row 2 column 2/.style={string replace={yes}{\cmark}},
  every row 3 column 2/.style={string replace={yes}{\cmark}},
  every row 4 column 1/.style={string replace={yes}{\cmark}},
  every row 5 column 1/.style={string replace={yes}{\cmark}},
  every row 4 column 2/.style={string replace={no}{\xmark}},
  every row 5 column 2/.style={string replace={no}{\xmark}},
  every row 6 column 1/.style={string replace={yes}{\cmark}},
  every row 7 column 1/.style={string replace={yes}{\cmark}},
  every row 6 column 2/.style={string replace={yes}{\cmark}},
  every row 7 column 2/.style={string replace={yes}{\cmark}},
  every head row/.style={before row=\toprule, after row=\midrule},
  every last row/.style={after row=\bottomrule},
  every row 0 column 0/.style={string replace={BEATs iter3+}{$\text{BEATs}_{\substack{\text{iter3+}}}$}},
  every row 1 column 0/.style={string replace={BEATs iter3+ ft}{$\text{BEATs}_{\substack{\text{iter3+}}}$ ft}},
  every row 2 column 0/.style={string replace={BEATs iter3+}{$\text{BEATs}_{\substack{\text{iter3+}}}$}},
  every row 3 column 0/.style={string replace={BEATs iter3+ ft}{$\text{BEATs}_{\substack{\text{iter3+}}}$ ft}},
  every row 4 column 0/.style={string replace={BEATs iter3+}{$\text{BEATs}_{\substack{\text{iter3+}}}$}},
  every row 5 column 0/.style={string replace={BEATs iter3+ ft }{$\text{BEATs}_{\substack{\text{iter3+}}}$ ft}},
  every row 6 column 0/.style={string replace={BEATs iter3+}{$\text{BEATs}_{\substack{\text{iter3+}}}$}},
  every row 7 column 0/.style={string replace={BEATs iter3+ ft }{$\text{BEATs}_{\substack{\text{iter3+}}}$ ft}},
]\datatable
\end{adjustbox}
\end{table}

\pgfplotstableread[col sep=comma]{data/table3_pooling2.csv}\datatable
\begin{table}[t]
\centering
\caption{Comparison of pooling methods.}
\label{tab:result_pooling}
\begin{adjustbox}{width=0.44\textwidth}
\pgfplotstabletypeset[
  col sep=comma,
  string type,
  columns/Method/.style={string type, column type=l, column name={Method}},
  columns/DimFeature/.style={string type, column type=l, column name={$C$}},
  columns/auc2020/.style={string type, column name=AUC, column type=l},
  columns/pauc2020/.style={string type, column name=pAUC, column type=l},
  columns/auc2023/.style={column name=AUC, column type=c, precision=2, fixed zerofill},
  columns/pauc2023/.style={column name=pAUC, column type=c, precision=2, fixed zerofill},
  every col no 0/.style={string type, column type=l, column name=Method},
  every col no 1/.style={string type, column type=c, column name={Dim $C$}},
  every last row/.style={after row=\bottomrule},
    every head row/.style={
    before row={
      \toprule
      \multicolumn{2}{c}{} & \multicolumn{2}{c}{DCASE2020} & \multicolumn{2}{c}{DCASE2023} \\
    },
    after row=\midrule
  },
]\datatable
\end{adjustbox}
\end{table}

\subsubsection{Comparison to state-of-the-art}
\label{sssec:comparesota}
The results for domain-generalized ASD on DCASE2023T2 Eval set are shown in Table \ref{tab:result_official}. We report results for \textit{GenRep} with different values of $K_s$ (100, 500, and 990), with the best performance observed at $K_s = 990$. Our methods outperform the state-of-the-art (SOTA) OE performance across all main metrics: AUC source (81.30\% vs. 75.50\%), AUC target (77.51\% vs. 68.70\%), and pAUC (64.71\% vs. 61.60\%). The primary performance gain of our approach over existing methods is the significantly higher AUC score on the target domain, achieving a higher score than the SOTA OE by 8.81\%. Our method is robust across different machine types, with the lowest AUC score being 71.92\% on the ToyTank dataset, while the highest performance is observed on the Vacuum dataset with an AUC score of 96.77\%. Overall, our method demonstrates robustness under domain shift settings with limited availability of target samples, achieving an official score of 73.79.

\pgfplotstableread[col sep=comma]{data/table_low-shot3.csv}\datatable
\begin{table}[t]
\centering
\caption{Result of low-shot experiment on DCASE2020T2 Dev set \cite{koizumi_description_2020}.}
\label{tab:low-shot}
\begin{adjustbox}{width=0.43\textwidth}
\pgfplotstabletypeset[
  col sep=comma,
  string type,
  every col no 0/.style={string type, column type=l, column name=Method},
  every head row/.style={before row=\toprule, after row=\midrule},
  every row no 6/.style={after row=\midrule},
  every row no 7/.style={before row=\rowcolor[gray]{0.9}},
  every last row/.style={after row=\bottomrule},
  every row 8 column 0/.style={string replace={AE}{AE \cite{koizumi_description_2020}}},
  every row 9 column 0/.style={string replace={AE-IDNN}{AE-IDNN \cite{suefusa_anomalous_2020}}},
  every row 10 column 0/.style={string replace={ANP-IDNN}{ANP-IDNN \cite{wichern_anomalous_2021}}},
  every row 11 column 0/.style={string replace={PAE}{PAE \cite{10095568}}},
  every row 12 column 0/.style={string replace={AudDSR}{AudDSR \cite{10447941}}},
  every row 13 column 0/.style={string replace={RSMM-MR}{RSMM-MR \cite{lee2021robust}}},
]\datatable
\end{adjustbox}
\end{table}

\subsection{Ablation studies}
Table \ref{tab:domainnorm} highlights the impact of DN and MemMixup ($K_s = 990$), in mitigating domain shift using $\text{BEATs}_{\substack{\text{iter3+}}}$ and $\text{BEATs}_{\substack{\text{iter3+}}}$ ft. The results show that these techniques balance performance between source and target domains. For $\text{BEATs}_{\substack{\text{iter3+}}}$ ft, DN alone boosts the target domain AUC by 16.61\%, while MemMixup alone improves it by 8.09\%. Although these methods enhance the target memory bank, they can reduce the source domain AUC due to test source samples becoming closer to the target memory bank. DN generally causes a larger drop in source AUC but yields a higher target AUC, improving the official score. Combining DN with MemMixup achieves the best performance across both domains, resulting in the highest official score.

To evaluate whether the temporal pooling is the most effective pooling method, we compare it with the spatial pooling and the spectral pooling, on the DCASE2020T2 Dev set and the DCASE2023T2 Eval set. As shown in Table \ref{tab:result_pooling},  temporal pooling consistently yielded better performance across all datasets. This suggests that preserving the spectral component is essential, as temporal pooling effectively removes temporal information while maintaining crucial spectral details.

\subsection{Robustness to limited data}
\label{ssec:fewshot}
We evaluate \textit{GenRep} in a low-shot setting with training samples ranging from 4 (0.12\% of the dataset) to 200 (5.96\%), using five different random seeds for the training split (except for the full-shot setting), and report the average scores. Note that MemMixup is not applied in this setting. Table \ref{tab:low-shot} shows the results of \textit{GenRep} compared to unsupervised generative approaches. Our observations are as follows: (1) With about 6\% of the training data (200-shot), \textit{GenRep} matches the performance of the AE baseline \cite{koizumi_description_2020}; (2) \textit{GenRep} in half-shot and full-shot settings is competitive with advanced AE approaches \cite{suefusa_anomalous_2020}\cite{wichern_anomalous_2021}\cite{10095568}. Furthermore, \textit{GenRep} can leverage section ID labels for domain normalization (DN), improving performance over the label-free method by up to 2.1\% (full-shot) in AUC scores.

\section{Conclusion}
\label{sec:conclusion}
In conclusion, our method, \textit{GenRep}, demonstrates strong performance and generalizability across both source and target domains, as well as in limited data scenarios. By leveraging BEATs \cite{pmlr-v202-chen23ag} as a pre-trained feature extractor without fine-tuning, along with techniques such as temporal pooling, MemMixup, and Domain Normalization, \textit{GenRep} effectively enhances domain-generalized ASD. This approach addresses key challenges, including domain shifts and noisy inputs, without the need for extensive labeled data or complex tuning, highlighting a promising direction for developing efficient and adaptable ASD systems.



\pagebreak
\begingroup
\ninept
\bibliographystyle{IEEEbib}
\bibliography{references, refs}
\endgroup

\end{document}